\pdfoutput=1
\documentclass[11pt]{article}

\usepackage{emnlp2021}
\usepackage{times}
\usepackage{latexsym}
\usepackage[T1]{fontenc}
\usepackage[utf8]{inputenc}
\usepackage{graphicx}
\usepackage{subcaption}
\usepackage{microtype}
\usepackage{xcolor}
\usepackage{amsmath}
\usepackage{multirow}
\usepackage{tabularx}
\usepackage{tikz}
\usepackage{pgfplots}
\usepackage{comment}
\usepackage{fancyref}
\usepackage{float}
\pgfplotsset{compat=1.7}

\title{Low Frequency Names Exhibit Bias and Overfitting in Contextualizing Language Models}

\author{Robert Wolfe \\
  University of Washington \\
  \texttt{rwolfe3@uw.edu} \\\And
  Aylin Caliskan\\
  University of Washington\\

  \texttt{aylin@uw.edu} \\}

\begin{document}
\maketitle

\begin{abstract}
We use a dataset of U.S. first names with labels based on predominant gender and racial group to examine the effect of training corpus frequency on tokenization, contextualization, similarity to initial representation, and bias in BERT, GPT-2, T5, and XLNet. We show that predominantly female and non-white names are less frequent in the training corpora of these four language models. We find that infrequent names are more self-similar across contexts, with Spearman's $\rho$ between frequency and self-similarity as low as $-.763$. Infrequent names are also less similar to initial representation, with Spearman's $\rho$ between frequency and linear centered kernel alignment (CKA) similarity to initial representation as high as $.702$. Moreover, we find Spearman's $\rho$ between racial bias and name frequency in BERT of $.492$, indicating that lower-frequency minority group names are more associated with unpleasantness. Representations of infrequent names undergo more processing, but are more self-similar, indicating that models rely on less context-informed representations of uncommon and minority names which are overfit to a lower number of observed contexts.
\end{abstract}

\section{Introduction}

Human social perception is linked to frequency of observation.  \citet{hughes2019neural} show using functional magnetic resonance imaging (fMRI) scans that humans are more aware of variation in the faces of members of their own race, and perceive members of other races as repeated instances of a social class, rather than as individuals. Most people interact more with individuals of their own race, and develop better cognitive skills for differentiating members of the race they see most frequently. Recent research indicates that state-of-the-art AI systems mirror such biased and unequal human perceptions. For example, \citet{pmlr-v81-buolamwini18a} show that under-representation in the training data of computer vision models causes poor performance on classification tasks for women and minority racial groups.

First names are used in both social psychology and Natural Language Processing (NLP) as a proxy to ground truth data for studying racial and gender biases. The implicit association test (IAT) of \citet{greenwald1998measuring} finds that study participants perceive European-American names as more pleasant than African-American names, and \citet{caliskan2017semantics} demonstrate with the word embedding association test (WEAT) that human biases observed in the IAT exist in static word embeddings, learned vector representations of words widely used in NLP. We use a list of first names labeled by gender and racial group based on U.S. Social Security Administration data and the names dataset of \citet{tzioumis2018demographic} to analyze how name frequency affects minority social groups in four neural language models: BERT, GPT-2, XLNet, and T5. Neural language models have advanced the state of the art in NLP, and are found in consequential NLP contexts such as Google Search \cite{nayak_2019}. These models produce contextualized word embeddings, which incorporate information from the context in which the word occurs over a series of neural network layers. \citet{may-etal-2019-measuring} and \citet{guodetecting} show that racial, gender, and intersectional biases exist in neural language models. We examine whether under-representation in the training corpora of such models causes them to overfit non-white and female names, reducing model generalization for underrepresented minorities.  We list our research questions and contributions:

\noindent (1) \textbf{Are minority social group names less frequent in the training corpora of neural language models?} We process four training corpora and find that white and male names are the most frequent in all training corpora. \\
\noindent (2) \textbf{Are infrequent and minority group names split into subwords by neural language models more than frequent and majority group names?} We report the percentage of names singly tokenized for eight demographic groups. Minority and female group names are singly tokenized less than white and male names. Single tokenization correlates with frequency, with Spearman's $\rho$ up to .835.  \\
\noindent (3) \textbf{Are infrequent minority group names more biased than frequent minority group names?} We take Spearman's $\rho$ of name frequency and bias using the WEAT. We find that infrequent racial minority group names are more negative in BERT, with Spearman's $\rho$ of frequency and pleasantness association of .492. Common female names exhibit greater gender bias in BERT, with Spearman's $\rho$ of frequency and career/family bias of -.553.\\
\noindent (4) \textbf{Are infrequent and minority group names less contextualized than frequent and majority group names? Is this the result of overfitting, or of underfitting?} We examine intra-layer self-similarity of embeddings across contexts to determine how contextualized a name is, and measure its similarity to initial representation in the model's embedding lookup matrix using linear centered kernel alignment (CKA) of \citet{kornblith2019similarity}. Inter-layer similarity to initial representation shows how much processing a name undergoes, indicating whether a poorly contextualized word is overfit to few observed contexts, if it's notably different from initial representation, or underfit, if it's similar to the representation in the embedding lookup matrix. We find that infrequent and minority group names exhibit higher intra-layer self-similarity, with Spearman's $\rho$ of frequency and intra-layer self-similarity as low as -.763. Infrequent and minority group names are less similar to initial representation, with Spearman's $\rho$ between name frequency and similarity to initial representation up to .702, suggesting overfitting.

We use Spearman's $\rho$ rather than Pearson's $\rho$ to capture monotonic with name frequency, as we observe effects primarily on a log scale. The null hypothesis is that frequency does not affect tokenization, social bias, and contextualization, and we disprove it by obtaining a two-tailed p-value. 
\section{Related Work}
We survey prior work related to frequency's influence on static and contextualized word embeddings. Static word embeddings are vector representations of words based on co-occurrence statistics of words in a training corpus. Contextualized word embeddings are vector representations of words from a neural language model which incorporates information from context to minimize loss on a training objective, such as next-word prediction (language modeling). Static word embeddings have one vector per word, while contextualized embeddings vary with context, allowing them to capture, for example, the sense of a polysemous word.

The WEAT of \citet{caliskan2017semantics} shows that the strength of association of a set of target static word embeddings (\textit{e.g.}, two social groups) to two sets of polar attribute static word embeddings (\textit{e.g.}, pleasant/unpleasant) encodes widely shared non-social biases, stereotypes, and factual information about the world.  \citet{may-etal-2019-measuring} extend the WEAT to neural language models with the Sentence Encoder Association Test (SEAT), which inserts WEAT target words into semantically bleached sentence templates such as "This is <word>" and measures the association of sentence vectors rather than word vectors. \citet{kurita-etal-2019-measuring} mask target and attribute words in template sentences, and directly query BERT’s masked language modeling objective to compute an association of target words to attributes. \citet{guodetecting} evaluate the overall magnitude of bias in language models by extending the WEAT to the Contextualized Embedding Association Test (CEAT).

\citet{brunet2019understanding} perturb the sparse word co-occurrence matrix of the GloVe static word embedding algorithm of \citet{pennington-etal-2014-glove} by omitting co-occurrence information from a particular subsection of the corpus, and show that
embeddings of rare words are the most biased and the most sensitive to corpus perturbations. \citet{wang2020doublehard} find that the frequency of a word in a training corpus can twist gender direction and create features in static word embeddings which vary based on frequency. Our work analyzes neural language models, which do not form representations directly based on co-occurrence statistics.

\citet{NEURIPS2018_e555ebe0} examine static Word2Vec and Transformer word embeddings used in machine translation and find that embeddings of high and low-frequency words lie in separate subregions of the embedding space, and that an embedding of a rare word and a common word can be distant even if they are semantically similar. The authors train an adversarial model to produce embeddings not separable based on frequency. \citet {provilkov-etal-2020-bpe} propose byte-pair encoding dropout, a subword regularization algorithm which allows multiple byte-pair encoding segmentations of the same word and corrects the problem of rare subtokens existing in a separate subregion of embedding space.

\citet{mu2018allbutthetop} show that the top two directions using Principal Component Analysis (PCA) in Skip-Gram, GloVe, and Continuous Bag of Words (CBOW) static word embeddings encode frequency-related features, and that removing frequency-related features improves performance on tasks related to word similarity. \citet{ott2018analyzing} find that beam search in neural machine translation prefers common tokens, including common subword tokens, to uncommon tokens. \citet{wendlandt-etal-2018-factors} measure the stability of static word embeddings, defined as consistency in the percent overlap of nearest neighbors, and find that frequency contributes to semantic stability in word embeddings.

\citet{ethayarajh-2019-contextual} demonstrate that contextualized word embeddings in BERT, GPT-2, and ELMo occupy an increasingly anisotropic vector space as they incorporate context. The most context-specific, least self-similar tokens are the most frequent in the training corpora, and in BERT, a token embedding becomes more similar to the embeddings in the context around it as it is contextualized. \citet{zhou2021frequencybased} find that frequency distorts the geometry of contextualized embeddings from BERT, which causes the model to over or underestimate semantic similarity of words based on frequency in the training corpus.

\section{Background: Neural Language Models}

Our work considers BERT, GPT-2, T5, and XLNet. We choose these neural language models both because they are commonly used and studied, and because they use three different subword tokenizers, allowing us to conduct our tokenization analysis across three different algorithms. We use the 12-layer cased implementation of each transformer in Python with TensorFlow and the Hugging Face Transformers library of \citet{wolf-etal-2020-transformers}. Appendix \ref{nlm_appendix} provides details on these models.

\subsection{Subword Tokenization}
Most neural language models use subword tokens to represent text. Each subword is tied to a vector in the model's embedding lookup matrix, which is trained with the model. Some common words in the training corpus are represented with a single embedding, but most are broken into subcomponents and mapped to multiple subword embeddings. Subword tokenization solves the out-of-vocabulary (OOV) problem, which occurs when a language model encounters a word not in its vocabulary. Subword tokenization allows a model to maintain a smaller vocabulary than a model with a full-word vocabulary, like TransformerXL of \citet{dai-etal-2019-transformer}, which has a vocabulary size of over 267,000, and uses the OOV token for words not in its vocabulary. Subword tokenization is faster than character convolutions to form an initial embedding, as in \citet{peters-etal-2018-deep} in ELMo, or \citet{boukkouri2020characterbert} in Character-BERT. All four language models examined use subword tokenization. BERT uses WordPiece, GPT-2 uses Byte-Pair Encoding, and XLNet and T5 use SentencePiece. Appendix \ref{tokenizer_appendix} provides further details.

\subsection{Representational Similarity Measures} 
\citet{voita-etal-2019-bottom} use projection-weighted canonical correlation analysis (PWCCA) to measure embedding change from layer to layer of language models trained for different tasks, a method developed by \citet{NEURIPS2018_a7a3d70c} which can measure the evolution of neural network representations. More recently, \citet{kornblith2019similarity} measure layer differences using linear CKA.

\citet{kiela2015exploiting} use dispersion of image vectors as a measure of the generality of an associated word to distinguish hypernyms from hyponyms. \citet{ethayarajh-2019-contextual} use self-similarity to measure contextualization in neural language models.

\subsection{Contextualized Embedding Extraction}

\citet{bommasani-etal-2020-interpreting} note that intermediate layers of neural language models are often used in downstream applications, and \citet{ethayarajh-2019-contextual} finds that static embeddings formed from the first two layers of BERT and GPT-2 are accurate for several common NLP tasks. \citet{DBLP:journals/corr/abs-1905-05950} show that layers of BERT attend primarily to a certain NLP task (such as coreference), and occur in an expected order. \citet{voita-etal-2019-bottom} find that a model pretrained for language modeling (next-word prediction), such as GPT-2, loses information about the current token while forming a prediction about the future, meaning that the top layer is poorly suited to analysis of the input token. 

We use the ValNorm method of \citet{toney2020valnorm} to choose the layer of each model which reflects the semantics of the input token. ValNorm obtains a valence score for each non-social group word in a lexicon by calculating its association with two sets of pleasant and unpleasant words based on cosine similarity. It then obtains Pearson's $\rho$ to measure the similarity of the valence scores of the word embeddings with a set of human-evaluated ratings. This is referred to as the ValNorm score, which the authors find is stable across languages, historical periods, and word embedding algorithms. ValNorm was developed for static word embeddings, and we extend it to measure the semantic quality of contextualized word embeddings in each layer of a neural language model.
\section{Datasets}
\label{sec:dataset}

We use a dataset of first names from \citet{tzioumis2018demographic} cross-referenced with U.S. Social Security Administration (SSA) data to analyze frequency in language models. To obtain contextualized representations of the names from language models, we gather contexts from the Reddit corpus of \citet{baumgartner2020pushshift}. For name frequency statistics, we process the training corpus of each model.

\begin{table*}[htbp]
\centering
\begin{tabular}
{|l||r|r|r|r|r|r|r|r|}
 \hline
 \multicolumn{9}{|c|}{Median Frequency of Names by Demographic Group} \\
 \hline

Training Corpus & \multicolumn{1}{c|}{AF} & \multicolumn{1}{|c|}{BF} & \multicolumn{1}{|c|}{HF} & \multicolumn{1}{|c|}{WF} & \multicolumn{1}{|c|}{AM} & \multicolumn{1}{|c|}{BM} & \multicolumn{1}{|c|}{HM} & \multicolumn{1}{|c|}{WM}\\
 \hline
   OpenWebText & 1,458 & 1,034 & 679 & 2,142 & 4,172 & 12,138 & 2,215 & 13,593\\
 BookCorpus & 378 & 578 & 219 & 1,191 & 483 & 2,644 & 274 & 3,074\\
Wikipedia & 5,660 & 1,566 & 3,372 & 5,968 & 10,102 & 21,857 & 8,384 & 28,455\\
C4 & 32,257 & 16,843 & 13,149 & 41,710 & 66,661 & 125,743 & 30,760 & 163,014\\
 \hline
\end{tabular}
\caption{White names and male names are the most common in reconstructions of training corpora.} 
\label{Training_Corpora_Table}
\end{table*}

\subsection{First Names} \label{subsection_first_names}
We obtain a dataset of first names segmented by demographic group by cross-referencing a list developed by \citet{tzioumis2018demographic} of first names labeled by racial self-identification with 1990 SSA data, which includes information about the gender and the frequency of a first name based on births in 1990. Tzioumis’ list combines data from three mortgage datasets, and Tzioumis estimates that it covers 85.6\% of the U.S. population, based on U.S. census data. We assign each name a label corresponding to the race as which the highest number of individuals possessing that name self-identify. Each name is assigned a gender based on the gender with the greater number of births with the name in the 1990 SSA data. The resulting dataset of 3,757 names includes two genders (Female, Male) and four racial groups (Asian, Black, Hispanic, White). Appendix \ref{first_names_appendix} includes more information.

\subsection{Contexts}

To measure contextualization, we extract contextualized embeddings from a variety of contexts. Following \citet{guodetecting}, we harvest contexts from the Reddit corpus of \citet{baumgartner2020pushshift} at \url{pushshift.io} . We remove contexts in which a name exists in a first-last name combination which refers to a public figure (\textit{e.g.,}, "Taylor Swift"), which could skew our analysis.

To control for the influence of context, we harvest $1,000$ contexts for the relatively unisex name "Taylor" (7,258 female births vs. 6,577 male births in 1990 SSA data), and create a set of identical contexts for every name in our dataset with each name replacing "Taylor." For example, we change “I saw Taylor last week” to “I saw Latisha last week” to represent a black woman.

\subsection{Training Corpora}

We obtain ground truth data for name frequency in the training corpora of each language model. Most of the training corpora are not available to the public, but can be replicated or approximated.

BERT was trained on the BookCorpus and English Wikipedia. The BookCorpus is no longer available, and we use the open source reproduction of \citet{soskkobayashi2018bookcorpus}. We use the September 20, 2017 English Wikipedia dump made available by Wikimedia\footnote{\url{archive.org/details/enwiki-20170920}}, from the time period BERT was being trained. We combine name counts from the BookCorpus and Wikipedia to obtain name frequencies for BERT and XLNet. XLNet is trained on three additional corpora which are available for a fee or which would be difficult to reconstruct. The .792 Spearman coefficient we obtain from combined BookCorpus and Wikipedia frequencies and single tokenization of names by XLNet indicates that these frequencies are representative of the entire XLNet training corpus.
The WebText corpus on which GPT-2 was trained was not made available to the public. We use the open source replication OpenWebText by \citet{Gokaslan2019OpenWeb}, produced based on the web crawling heuristic of \citet{radford2019language}.
T5 was trained on the Colossal Cleaned Crawled Corpus (C4). We obtain ground truth name frequencies for C4 from the 800GB cleaned version reconstructed by AllenAI\footnote{\url{github.com/allenai/allennlp/}}.

Table \ref{Training_Corpora_Table} shows that white male names have the highest median frequency in every training corpus, and the median frequency of any male group is higher than that of any female group in OpenWebText, Wikipedia, and C4. Black, Hispanic, and Asian females have the lowest median frequency in every training corpus.
\section{Approach and Experiments} 

We measure the effects of frequency on tokenization, social bias, and contextualization to understand disparities between majority social groups and the minority racial and gender social groups identifiable in our names dataset. 

\subsection{Tokenization}
We tokenize names in our dataset using the default Transformers library tokenizers of \citet{wolf-etal-2020-transformers} for BERT, GPT-2, T5, and XLNet. We obtain the proportion of names singly  tokenized for the racial and gender groups in our data. Tokenization by social group informs our other experiments, as contextualization varies based on tokenization.

\subsection{Bias}

We quantify the bias association of each name for five WEATs described by \citet{caliskan2017semantics}. These include pleasant/ unpleasant (25 words, 8 words), career/family, math/art, and science/art. Following WEAT, each set contains at least 8 words to satisfy concept representation significance. Accordingly, the limitations of not adhering to this methodological robustness rule of WEAT, which are outlined by \citet{ethayarajh2019understanding}, are mitigated. 
Pleasant/unpleasant tests measure racial bias, and career/family, math/art, and science/art tests measure gender bias. The formula for the single-value WEAT follows. $w$ refers to a target word, and $A$ and $B$ to sets of polar attribute words.
\[\frac{\textrm{mean}_{a\in A}\textrm{cos}(\vec{w},\vec{a}) - \textrm{mean}_{b\in B}\textrm{cos}(\vec{w},\vec{b})}{\textrm{std\_dev}_{x \in A \cup B}\textrm{cos}(\vec{w},\vec{x})}\]

We extract contextualized word embeddings using semantically bleached sentences as described by \citet{may-etal-2019-measuring}. For names, we use the semantically bleached sentence template “This person’s name is <name>.” For WEAT attribute words, we use the context “This is <word>.” Unlike \citet{may-etal-2019-measuring}, we extract a contextualized word embedding of the stimuli rather than a sentence vector.

The WEAT measures association using cosine similarity, which requires that vectors are of equal length, so we must choose a  pooling method to represent multiply subtokenized words. Following \citet{bommasani-etal-2020-interpreting}, who show that mean pooling produces accurate representations for word similarity tasks, we use the mean of subword vectors to represent multiply tokenized words.

We use the ValNorm method of \citet{toney2020valnorm} to select an intermediate layer from which to extract contextualized word embeddings. ValNorm evaluates semantic quality based on the single-value WEAT. To adapt ValNorm for language models, we pool single-value WEAT comparisons to obtain a combined effect size as described by \citet{guodetecting}, and select the layer which produces embeddings best corresponding to human evaluations of word valence. Because our work uses the WEAT to measure bias, ValNorm is an especially useful method for evaluating which layers encode semantic information related to the current token. With ValNorm, we obtain Pearson’s $\rho$ of .881 in BERT layer 9, .859 in GPT-2 layer 7, .892 in T5 encoder layer 12, and .854 in XLNet layer 5. We refer to the layer with the highest ValNorm score as the semantic layer.

We obtain bias scores in the semantic layer for each name using the SV-WEAT on each bias test, and take Spearman's $\rho$ between bias and frequency for each minority group name in five tests.

\subsection{Contextualization}
Contextualization measures how much information from context a word incorporates. Words that incorporate information from context generalize well in contextualizing language models, whereas words which do not incorporate contextual information generalize poorly. To quantify contextualization, we measure intra-layer self-similarity across contexts. If a word is less contextualized, we seek to understand whether this is due to overfitting, or underfitting. How much processing a word undergoes in a model can help to indicate whether it is overfit or underfit. If the model changes the word significantly from initial representation, but fails to contextualize it, this suggests that the word is overfit, and processed to be similar to its embedding in few other contexts. If the model does not change the word much from its initial representation, and fails to contextualize it, this suggests that the model relies on a general representation, similar to what it sees in its embedding lookup matrix, indicating that the representation may be underfit. We measure inter-layer self-similarity to initial representation to determine whether poorly contextualized words are overfit or underfit.

For contextualization, we use concatenations of subword vectors for multiply tokenized words, rather than mean pooling, to compare representations closest to the way the model sees them. 

\subsubsection{Intra-Layer Self-Similarity}
We measure intra-layer self-similarity for each name in 1,000 identical contexts. Intra-layer self-similarity is the mean cosine similarity of contextualized embeddings for a word generated in a single layer of a language model, and  ranges between 0 (dissimilar) and 1 (similar). The formula below was described by \citet{ethayarajh-2019-contextual}, but removes a function to map a word to a layer and sentence index.

\[s(w) = \frac{1}{n^2 - n} \sum_{i} \sum_{j \neq i}  cos(\vec{w_i}, \vec{w_j})
\]

We take Spearman's $\rho$ of frequency and intra-layer self-similarity, and compare mean self-similarity for singly and multiply tokenized names.

\subsubsection{Inter-Layer Similarity}
We use linear CKA, a similarity index ranging between 0 (dissimilar) and 1 (similar), to measure similarity across layers. Proposed by \citet{kornblith2019similarity} for measuring similarity in neural network representations, linear CKA is not invariant to invertible linear transformation, but is invariant to orthogonal transformation and isotropic scaling, and identifies correspondences between layers more accurately than PWCCA. Invariance to isotropic scaling is useful, as \citet{ethayarajh-2019-contextual} found that upper layers of BERT and GPT-2 exhibit high anisotropy. Linear CKA operates on matrices of observations. We form matrices of 1,000 contextualized word embeddings for each name, inserted into the same 1,000 contexts as every other name. Below is the formula for linear CKA.

\begin{equation*}
\frac{||Y^TX||\genfrac{}{}{0pt}{2}{2}{\textrm{F}}}{||X^TX||{\genfrac{}{}{0pt}{2}{}{\textrm{F}}}||Y^TY||\genfrac{}{}{0pt}{2}{}{\textrm{F}}}
\end{equation*}

$X$ and $Y$ are matrices of examples, and $||\cdot|| {\genfrac{}{}{0pt}{2}{}{\textrm{F}}}$ refers to the Froebenius norm. We report Spearman's $\rho$ of name frequency and CKA similarity, and compare mean CKA similarity for singly and multiply tokenized names. 
\section{Results}
We find that common names are singly tokenized more than less common names; that name frequency correlates with social bias in BERT, and to a lesser extent in GPT-2 and T5; and that infrequent and multiply tokenized names exhibit higher intra-layer self-similarity and  lower inter-layer similarity to initial representation, suggesting overfitting.

\begin{figure}[htbp]
        \centering
        \includegraphics[height=2.2in]{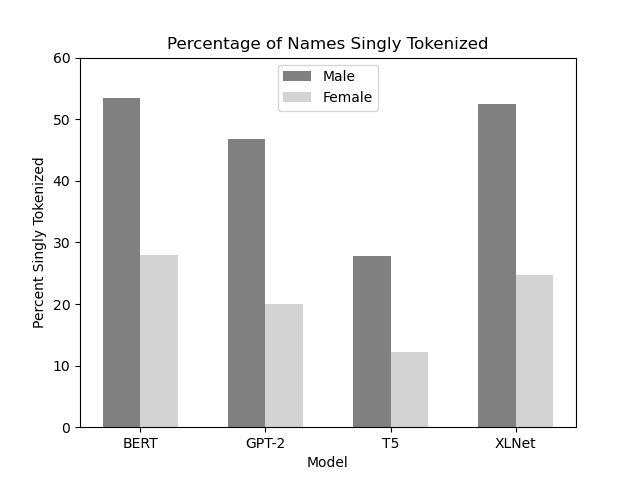}
        \caption{Male vs. female name tokenization}
    \label{tokenization_figure}
\end{figure}

\subsection{Tokenization}

Common names are singly tokenized more than uncommon names. Spearman's $\rho$ between frequency and single tokenization is .835 for BERT, .772 for GPT-2, .630 for T5, and .792 for XLNet. Commensurate with these correlations, Table \ref{Tokenization_Intersectional_Table} shows that white male names are the most singly tokenized in each language model. In BERT and XLNet, male names are singly tokenized more than female names, and white names are singly tokenized more than Asian, black, and Hispanic names.

\begin{table*}[htbp]
\centering
\begin{tabular}
{|l||r|r|r|r|r|r|r|r|}
 \hline
 \multicolumn{9}{|c|}{Proportion of Singly Tokenized Words by Demographic Group} \\
 \hline

Model & \multicolumn{1}{|c|}{AF} & \multicolumn{1}{|c|}{BF} & \multicolumn{1}{|c|}{HF} & \multicolumn{1}{|c|}{WF} & \multicolumn{1}{|c|}{AM} & \multicolumn{1}{|c|}{BM} & \multicolumn{1}{|c|}{HM} & \multicolumn{1}{|c|}{WM}\\
 \hline
   BERT & .218 & .081 & .122 & .308 & .331 & .563 & .329 & .619\\
 GPT-2 & .250 & .541 & .085 & .211 & .376 & .438 & .243 & .535\\
T5 & .199 & .027 & .074 & .122 & .198 & .219 & .119 & .328\\
XLNet & .308 & .162 & .079 & .262 & .392 & .531 & .337 & .590\\
 \hline
\end{tabular}
\caption{White male names are most singly tokenized.} 
\label{Tokenization_Intersectional_Table}
\end{table*}

\subsection{Bias}

Table \ref{Bias_Correlation_Table} shows Spearman's $\rho$ of bias score and training corpus frequency for minority group names. Pleasant/unpleasant (PU) coefficients measure correlation between frequency and association with pleasantness for non-white names.  Career/family (CF), math/art (MA), and science/art (SA) measure correlation between frequency and association with gender stereotypes for female names. 

\begin{table}[htbp]
\resizebox{\columnwidth}{!}{
\centering
\begin{tabular}
{|l||r|r|r|r|}
 \hline
 \multicolumn{5}{|c|}{Correlation of Bias and Frequency} \\
 \hline
 Bias Test & \multicolumn{1}{c}{BERT} & \multicolumn{1}{|c|}{GPT-2} & \multicolumn{1}{|c|}{T5} & \multicolumn{1}{|c|}{XLNet}\\
 \hline
 Race PU25 & .492 & -.011 & .020 & -.139 \\
 Race PU8 & -.021 & -.022 & .229 & .020 \\
 Gender CF & -.553 & .065 & .063 & -.333 \\
 Gender MA & -.311 & .139 & .094 & -.199 \\
 Gender SA & -.304 & .244 & -.080 & .164 \\
 \hline
\end{tabular}
}
\caption{Infrequent non-white names are more negative in BERT and T5.}
\label{Bias_Correlation_Table}
\end{table}

BERT and T5 exhibit positive correlation between pleasantness and minority group name frequency, with Spearman's $\rho$ of $.492$ and p-value of $10^{-167}$ for BERT. For these two models, more frequent minority group names are more associated with pleasantness. BERT also exhibits negative correlations between gender bias and frequency, indicating that more frequent observation of a female name reinforces its association with female stereotypes. The correlation between frequency in the training corpus and association with career as opposed to family in BERT is $-.553$ for female names, with a p-value of $10^{-161}$. XLNet exhibits similar correlations for the career/family and math/arts WEATs. P-values for correlations greater than $0.1$ or less than $-0.1$ are less than $10^{-10}$.

\subsection{Contextualization}

Infrequent names exhibit higher intra-layer self-similarity in each examined layer of each language model, and exhibit lower inter-layer similarity to initial representation, indicating that infrequent names generalize poorly to context and are overfit to the contexts in which they have been observed.

\subsubsection{Intra-Layer Similarity}

Table \ref{Frequency_Selfsim_Table} shows negative Spearman's $\rho$ in BERT, GPT-2, T5, and XLNet, ranging between $-.523$ and $-.763$, with p-values $<10^{-263}$. Most of the least negative correlations occur in the top layer, possibly due to high anisotropy in upper layers of language models observed by \citet{ethayarajh-2019-contextual}. Correlation of frequency and intra-layer self-similarity is highest in the first layer of GPT-2 and XLNet.

\begin{table}[htbp]
\centering
\begin{tabular}
{|l||r|r|r|r|}
 \hline
 \multicolumn{5}{|c|}{Correlation of Frequency and Self-Similarity} \\
 \hline
 Layer & BERT & GPT-2 & T5 & XLNet\\
 \hline
   First & -.688 & -.703 & -.523 & -.763 \\
 Second & -.683 & -.681 & -.574 & -.732 \\
Semantic & -.693 & -.633 & -.573 & -.698 \\
Output & -.598 & -.640 & -.573 & -.649 \\
 \hline
\end{tabular}
\caption{Infrequent names exhibit higher intra-layer self-similarity, and are less contextualized.}
\label{Frequency_Selfsim_Table}
\end{table}

\begin{figure}[htbp]
    \centering
        \includegraphics[height=2.2in]{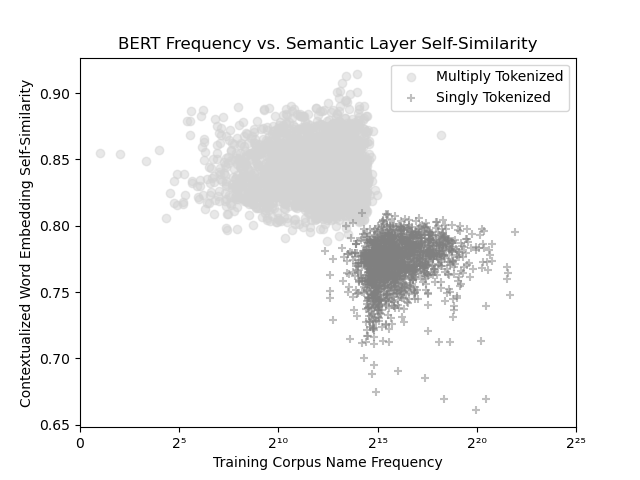}
    \caption{Multiply tokenized and infrequent names exhibit high intra-layer self-similarity in BERT.}
    \label{Frequency_Selfsim_Figure}
\end{figure}

Table \ref{Tokenization_Selfsim_Table} shows that multiply tokenized names exhibit higher intra-layer self-similarity than singly tokenized names. More common names, and names of majority group members, are more contextualized in language models than are infrequent names and names of minority group members.

\begin{table*}[htbp]
\centering
\begin{tabular}
{|l||r|r|r|r|r|r|r|r|}
 \hline
 \multicolumn{9}{|c|}{Mean Intra-Layer Self-Similarity by Tokenization} \\
 \hline
 \multirow{2}{*}{Layer} & \multicolumn{2}{c}{BERT} & \multicolumn{2}{|c|}{GPT-2} & \multicolumn{2}{|c|}{T5} & \multicolumn{2}{|c|}{XLNet}\\
 \cline{2-9}
 &\multicolumn{1}{|c|}{Single} & \multicolumn{1}{|c|}{Multi} & \multicolumn{1}{|c|}{Single} & \multicolumn{1}{|c|}{Multi} & \multicolumn{1}{|c|}{Single} & \multicolumn{1}{|c|}{Multi} & \multicolumn{1}{|c|}{Single} & \multicolumn{1}{|c|}{Multi}\\
 \hline
   First & .901 & .923 & .761 & .855 & .933 & .974 & .848 & .915\\
 Second & .863 & .926 & .762 & .815 & .916 & .965 & .759 & .846\\
Semantic & .772 & .840 & .736 & .826 & .702 & .769 & .613 & .653\\
Output & .771 & .808 & .810 & .974 & .702 & .769 & .863 & .926\\
 \hline
\end{tabular}
\caption{Multiply tokenized names exhibit higher intra-layer self-similarity.}
\label{Tokenization_Selfsim_Table}
\end{table*}

\subsubsection{Inter-Layer Similarity}

Table \ref{Frequency_CKA_Table} reports statistically significant positive correlations between frequency and inter-layer CKA similarity to initial representation, except in the first layer of GPT-2. Spearman's $\rho$ ranges between .174 and .702, with p-values smaller than $10^{-27}$. 

\begin{table}[htbp]
\centering
\begin{tabular}
{|l||r|r|r|r|}
 \hline
 \multicolumn{5}{|c|}{Correlation of Frequency and CKA Similarity} \\
 \hline
 Layer & BERT & GPT-2 & T5 & XLNet \\
 \hline
   First & .461 & -.299 & .174 & .623 \\
 Second & .592 & .487 & .382 & .636 \\
Semantic & .702 & .465 & .513 & .677 \\
Output & .480 & .242 & .513 & .338 \\
 \hline
\end{tabular}
\caption{Infrequent names exhibit lower inter-layer CKA  Similarity to initial representation.}
\label{Frequency_CKA_Table}
\end{table}

Embeddings of infrequent names are less similar to initial representation, indicating that poorly contextualized infrequent names are overfit to previously observed contexts, and that representations in the embedding lookup matrix carry more information for common and singly tokenized names.

\begin{table*}[htbp]
\centering
\begin{tabular}
{|l||r|r|r|r|r|r|r|r|}
 \hline
 \multicolumn{9}{|c|}{Mean CKA Similarity by Tokenization} \\
 \hline
 \multirow{2}{*}{Layer} & \multicolumn{2}{c}{BERT} & \multicolumn{2}{|c|}{GPT-2} & \multicolumn{2}{|c|}{T5} & \multicolumn{2}{|c|}{XLNet}\\
 \cline{2-9}
 &\multicolumn{1}{c|}{Single} & \multicolumn{1}{|c|}{Multi} & \multicolumn{1}{|c|}{Single} & \multicolumn{1}{|c|}{Multi} & \multicolumn{1}{|c|}{Single} & \multicolumn{1}{|c|}{Multi} & \multicolumn{1}{|c|}{Single} & \multicolumn{1}{|c|}{Multi}\\
 \hline
   First & .860 & .831 & .020 & .026 & .705 & .660 & .835 & .763\\
 Second & .736 & .633 & .013 & .006 & .564 & .475 & .674 & .515\\
Semantic & .149 & .088 & .004 & .001 & .398 & .201 & .370 & .238\\
Output & .064 & .051 & .016 & .015 & .398 & .201 & .002 & .001\\
 \hline
\end{tabular}
\caption{Multiply tokenized names are less similar to initial representation than singly tokenized names.}
\label{Tokenization_CKA_Table}
\end{table*}

Results for tokenization and mean inter-layer self-similarity are reported in Table \ref{Tokenization_CKA_Table}. We observe that contextualized representations of multiply tokenized names are less similar to their initial representations than singly tokenized names.  

\begin{figure}[htbp]
        \centering
        \includegraphics[height=2.2in]{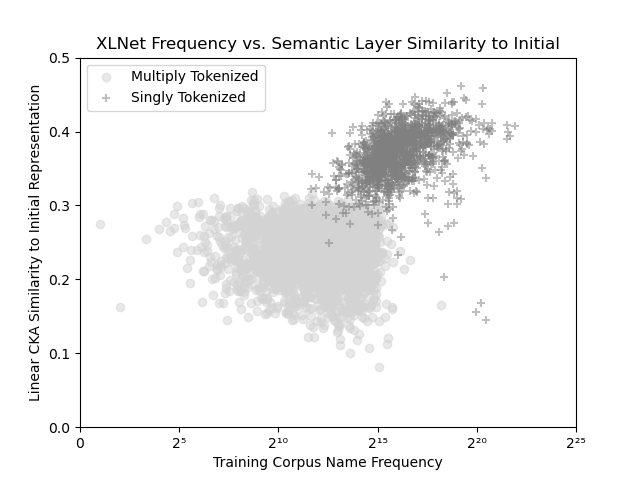}
    \caption{Infrequent names are less similar to initial representation in the semantic layer of XLNet.}
    \label{Frequency_CKA_Figure}
\end{figure}
\section{Discussion}

White male names occur more than any other social group in all training corpora we process. Black and Hispanic female names are least frequent, showing that frequency-based disparities in language models have notable consequences for non-white females. Each language model represents white male names with a single token more than any other group. Black and Hispanic female names are the least represented with a single subtoken, with less than 10\% singly tokenized in most models, reflecting disparities based on gender and race.

In BERT, we observe Spearman's $\rho$ of name frequency and racial bias of .492, indicating that less frequent minority names are more associated with unpleasantness. Gender bias in BERT exhibits negative correlation with frequency, indicating that it is reinforced by frequently observing a female name. While increased representation may address disparities in contextualization, fair and diverse context occurrence is also important to ensure that language models do not learn biased embeddings.

Infrequent and multiply tokenized names exhibit higher intra-layer self-similarity across language models, but lower inter-layer similarity to initial representation. Such names incorporate less information from context, and change more from initial representation, suggesting that disparities in contextualization are the result of overfitting.

Our work focuses on first names, but under-representation of minority names suggests that words and ideas specific to minority groups are also likely under-represented in the training corpora of neural language models, and will have the same issues with regard to tokenization, bias, and contextualization. Increased representation of minority groups in training corpora would mitigate many frequency-related problems, and correct biases in the corpora themselves. However, some words will always be more common than others. To reduce frequency-based disparities, language models trained on multiple objectives, such as T5, might incorporate an adversarial training objective such as the method described by \citet{NEURIPS2018_e555ebe0}.

Limitations of our work relate to using a list of names, which cannot capture many characteristics which may be biased in language models, such as sexual orientation. Please see our Ethical Considerations section for a detailed discussion of ethical concerns and limitations related to our work.
\section{Conclusion}

We show that names predominantly belonging to members of minority social groups occur less frequently than majority group names in training corpora of neural language models, and that low frequency results in lower rates of single tokenization, less contextualization, and overfit representations in neural language models.
\section{Ethical Considerations}

Our work with large-scale demographic datasets has involved simplifications which, while useful for studying the impact of language models on marginalized demographic groups, has also introduced a number of limitations to our work. These are primarily related to the use of categorical labels for gender and race, but also concern the properties of the demographic datasets employed and the role of frequency in interpreting our results.

A central limitation of our work is that categorical labels are assigned to each name based on gender and race. We understand race and gender not as essential characteristics, but as defined within a particular culture, time period, and social structure. Our research question is whether underrepresentation (\textit{i.e.,} low frequency) of social groups in a language model training corpus results in overfitting and exacerbated bias in the trained language model. To answer this, we analyze frequency of representation and its effects along two axes of demographic bias: gender bias, wherein males are more represented than females; and racial bias, wherein people considered to be white are represented more than people considered to be black, Hispanic, or Asian. The range of social groups studied reflects not an ideal division of humanity into immutable categories, but a comprehensible way of interpreting the disadvantages caused by underrepresentation in our current cultural context.
 
Our work can be thought of an exposure study which examines the effects of one aspect of gender and race: underrepresentation. In the context of studies of race, \citet{sen2016race} usefully define an exposure study as one which uses "a cue or signal that generates some reaction," and note that "names often act as a proxy for for traits associated with racial or ethnic groups." Well-established NLP methods such as the WEAT of \citet{caliskan2017semantics} have demonstrated the efficacy of using first names for observing biases in AI. As described in Section \ref{subsection_first_names}, we use U.S. Social Security Administration (SSA) data to give each first name a gender label, and a dataset which uses the same racial categories as the U.S. census does for surnames to give each first name a racial label. We do not intend to essentialize the name, or individuals bearing the name, but to identify the gender or racial association most likely to be perceived by a language model within the cultural context in which it was trained. We regret that many individuals are not represented by this set of categories. This is true for transgender and nonbinary individuals, as the SSA data reflects a gender binary, and also for people whose racial identity is not captured by one of a few categories.

Census data includes information about racial identification for surnames, but not for first names, prompting us to seek another source for this data. We choose the dataset of \citet{tzioumis2018demographic} because it has wide coverage (85.6\% of the U.S. population as estimated by Tzioumis); is ethically anonymized; is based on at least 30 observations for 91.2\% of included names; and contains first name data based on self-identification, considered a “gold standard” for demographic classification, as noted by \citet{larson2017gender} in the context of labeling data by gender. Tzioumis’ dataset uses the same racial categories as U.S. census data for surnames, indicating that these categories correspond to our cultural context.

That many names occur in more than one racial group or gender group is also a limitation, and may introduce noise into our analyses. However, more than 80\% of names examined in our work have a self-identification rate of at least 70\% with a single racial group in the Tzioumis dataset. Based on SSA data, 31.3\% of the names in our study have only male occurrences, 38.6\% of the names in our study have only female occurrences, and 30.1\% of names occur for both male and female individuals. 88.0\% of names with male and female occurrences have at least 70\% of occurrences associated with one gender. Thus, assigning a label based on the group with the most occurrences is likely to capture both which groups are most affected by our findings, as well as the linguistic signals related to race and gender which make first names a useful proxy to this information. 

We also note that the time periods from which demographic and language modeling data are derived may inject some noise into our analysis. We cross-reference SSA data on births in 1990 with the Tzioumis dataset, which is based on 2007 and 2010 mortgage applications. The populations studied in these two datasets are unlikely to overlap exactly, though they both sample the adult U.S. population during the time period germane to our research. Moreover, language models are themselves products of a specific place and time, as they are trained on text corpora assembled largely within a specific time period, and are unlikely to reflect the full diversity of language within that time period; OpenAI's WebText corpus, for example, is collected using outbound web links from Reddit running through December 2017 \cite{radford2019language}.

Finally, we note that our results are the effect primarily of low frequency of observation, and it is possible that similar differences in tokenization and contextualization based on frequency might be observed in other sets of words which are unrelated to demographics. However, frequency relates directly to the social dimension to which our work directs attention: that names of marginalized social groups are underrepresented - less frequent - in language modeling corpora, and more likely to be overfit by a language model. Frequency of representation is one of the composite variables that inform race and gender in our cultural context, and must be considered when training a language model which forms differing representations based on frequency. 

While large-scale demographic datasets are reductive, they can help to observe the social impact of the systems we study. Within the limited confines of the groups defined, our work shows that neural language models are ill-adapted to represent marginalized social groups.

\bibliography{anthology,custom}

\begin{thebibliography}{47}
\expandafter\ifx\csname natexlab\endcsname\relax\def\natexlab#1{#1}\fi

\bibitem[{Baumgartner et~al.(2020)Baumgartner, Zannettou, Keegan, Squire, and
  Blackburn}]{baumgartner2020pushshift}
Jason Baumgartner, Savvas Zannettou, Brian Keegan, Megan Squire, and Jeremy
  Blackburn. 2020.
\newblock The pushshift reddit dataset.
\newblock In \emph{Proceedings of the International AAAI Conference on Web and
  Social Media}, volume~14, pages 830--839.

\bibitem[{Bommasani et~al.(2020)Bommasani, Davis, and
  Cardie}]{bommasani-etal-2020-interpreting}
Rishi Bommasani, Kelly Davis, and Claire Cardie. 2020.
\newblock \href {https://doi.org/10.18653/v1/2020.acl-main.431} {{I}nterpreting
  {P}retrained {C}ontextualized {R}epresentations via {R}eductions to {S}tatic
  {E}mbeddings}.
\newblock In \emph{Proceedings of the 58th Annual Meeting of the Association
  for Computational Linguistics}, pages 4758--4781, Online. Association for
  Computational Linguistics.

\bibitem[{Boukkouri et~al.(2020)Boukkouri, Ferret, Lavergne, Noji, Zweigenbaum,
  and Tsujii}]{boukkouri2020characterbert}
Hicham~El Boukkouri, Olivier Ferret, Thomas Lavergne, Hiroshi Noji, Pierre
  Zweigenbaum, and Junichi Tsujii. 2020.
\newblock Characterbert: Reconciling elmo and bert for word-level
  open-vocabulary representations from characters.
\newblock \emph{arXiv preprint arXiv:2010.10392}.

\bibitem[{Brunet et~al.(2019)Brunet, Alkalay-Houlihan, Anderson, and
  Zemel}]{brunet2019understanding}
Marc-Etienne Brunet, Colleen Alkalay-Houlihan, Ashton Anderson, and Richard
  Zemel. 2019.
\newblock Understanding the origins of bias in word embeddings.
\newblock In \emph{International Conference on Machine Learning}, pages
  803--811. PMLR.

\bibitem[{Buolamwini and Gebru(2018)}]{pmlr-v81-buolamwini18a}
Joy Buolamwini and Timnit Gebru. 2018.
\newblock \href {http://proceedings.mlr.press/v81/buolamwini18a.html} {Gender
  shades: Intersectional accuracy disparities in commercial gender
  classification}.
\newblock In \emph{Proceedings of the 1st Conference on Fairness,
  Accountability and Transparency}, volume~81 of \emph{Proceedings of Machine
  Learning Research}, pages 77--91, New York, NY, USA. PMLR.

\bibitem[{Caliskan et~al.(2017)Caliskan, Bryson, and
  Narayanan}]{caliskan2017semantics}
Aylin Caliskan, Joanna~J Bryson, and Arvind Narayanan. 2017.
\newblock Semantics derived automatically from language corpora contain
  human-like biases.
\newblock \emph{Science}, 356(6334):183--186.

\bibitem[{Dai et~al.(2019)Dai, Yang, Yang, Carbonell, Le, and
  Salakhutdinov}]{dai-etal-2019-transformer}
Zihang Dai, Zhilin Yang, Yiming Yang, Jaime Carbonell, Quoc Le, and Ruslan
  Salakhutdinov. 2019.
\newblock \href {https://doi.org/10.18653/v1/P19-1285} {Transformer-{XL}:
  Attentive language models beyond a fixed-length context}.
\newblock In \emph{Proceedings of the 57th Annual Meeting of the Association
  for Computational Linguistics}, pages 2978--2988, Florence, Italy.
  Association for Computational Linguistics.

\bibitem[{Devlin et~al.(2019)Devlin, Chang, Lee, and
  Toutanova}]{devlin-etal-2019-bert}
Jacob Devlin, Ming-Wei Chang, Kenton Lee, and Kristina Toutanova. 2019.
\newblock \href {https://doi.org/10.18653/v1/N19-1423} {{BERT}: Pre-training of
  deep bidirectional transformers for language understanding}.
\newblock In \emph{Proceedings of the 2019 Conference of the North {A}merican
  Chapter of the Association for Computational Linguistics: Human Language
  Technologies, Volume 1 (Long and Short Papers)}, pages 4171--4186,
  Minneapolis, Minnesota. Association for Computational Linguistics.

\bibitem[{Ethayarajh(2019)}]{ethayarajh-2019-contextual}
Kawin Ethayarajh. 2019.
\newblock \href {https://doi.org/10.18653/v1/D19-1006} {How contextual are
  contextualized word representations? comparing the geometry of {BERT},
  {ELM}o, and {GPT}-2 embeddings}.
\newblock In \emph{Proceedings of the 2019 Conference on Empirical Methods in
  Natural Language Processing and the 9th International Joint Conference on
  Natural Language Processing (EMNLP-IJCNLP)}, pages 55--65, Hong Kong, China.
  Association for Computational Linguistics.

\bibitem[{Ethayarajh et~al.(2019)Ethayarajh, Duvenaud, and
  Hirst}]{ethayarajh2019understanding}
Kawin Ethayarajh, David Duvenaud, and Graeme Hirst. 2019.
\newblock Understanding undesirable word embedding associations.
\newblock \emph{arXiv preprint arXiv:1908.06361}.

\bibitem[{Gage(1994)}]{gage1994new}
Philip Gage. 1994.
\newblock A new algorithm for data compression.
\newblock \emph{C Users Journal}, 12(2):23--38.

\bibitem[{Gokaslan and Cohen(2019)}]{Gokaslan2019OpenWeb}
Aaron Gokaslan and Vanya Cohen. 2019.
\newblock Openwebtext corpus.
\newblock \url{http://Skylion007.github.io/OpenWebTextCorpus}.

\bibitem[{Gong et~al.(2018)Gong, He, Tan, Qin, Wang, and
  Liu}]{NEURIPS2018_e555ebe0}
Chengyue Gong, Di~He, Xu~Tan, Tao Qin, Liwei Wang, and Tie-Yan Liu. 2018.
\newblock \href
  {https://proceedings.neurips.cc/paper/2018/file/e555ebe0ce426f7f9b2bef0706315e0c-Paper.pdf}
  {Frage: Frequency-agnostic word representation}.
\newblock In \emph{Advances in Neural Information Processing Systems},
  volume~31. Curran Associates, Inc.

\bibitem[{Greenwald et~al.(1998)Greenwald, McGhee, and
  Schwartz}]{greenwald1998measuring}
Anthony~G Greenwald, Debbie~E McGhee, and Jordan~LK Schwartz. 1998.
\newblock Measuring individual differences in implicit cognition: the implicit
  association test.
\newblock \emph{Journal of personality and social psychology}, 74(6):1464.

\bibitem[{Guo and Caliskan(2021)}]{guodetecting}
Wei Guo and Aylin Caliskan. 2021.
\newblock Detecting emergent intersectional biases: Contextualized word
  embeddings contain a distribution of human-like biases.
\newblock In \emph{Proceedings of the 2021 AAAI/ACM Conference on AI, Ethics,
  and Society}.

\bibitem[{Hughes et~al.(2019)Hughes, Camp, Gomez, Natu, Grill-Spector, and
  Eberhardt}]{hughes2019neural}
Brent~L Hughes, Nicholas~P Camp, Jesse Gomez, Vaidehi~S Natu, Kalanit
  Grill-Spector, and Jennifer~L Eberhardt. 2019.
\newblock Neural adaptation to faces reveals racial outgroup homogeneity
  effects in early perception.
\newblock \emph{Proceedings of the National Academy of Sciences},
  116(29):14532--14537.

\bibitem[{Johnson et~al.(2017)Johnson, Schuster, Le, Krikun, Wu, Chen, Thorat,
  Vi{\'e}gas, Wattenberg, Corrado et~al.}]{johnson2017google}
Melvin Johnson, Mike Schuster, Quoc~V Le, Maxim Krikun, Yonghui Wu, Zhifeng
  Chen, Nikhil Thorat, Fernanda Vi{\'e}gas, Martin Wattenberg, Greg Corrado,
  et~al. 2017.
\newblock Google’s multilingual neural machine translation system: Enabling
  zero-shot translation.
\newblock \emph{Transactions of the Association for Computational Linguistics},
  5:339--351.

\bibitem[{Kiela et~al.(2015)Kiela, Rimell, Vulic, and
  Clark}]{kiela2015exploiting}
Douwe Kiela, Laura Rimell, Ivan Vulic, and Stephen Clark. 2015.
\newblock Exploiting image generality for lexical entailment detection.
\newblock In \emph{Proceedings of the 53rd Annual Meeting of the Association
  for Computational Linguistics (ACL 2015)}, pages 119--124. ACL; East
  Stroudsburg, PA.

\bibitem[{Kobayashi(2018)}]{soskkobayashi2018bookcorpus}
Sosuke Kobayashi. 2018.
\newblock Homemade bookcorpus.
\newblock \url{https://github.com/BIGBALLON/}.

\bibitem[{Kornblith et~al.(2019)Kornblith, Norouzi, Lee, and
  Hinton}]{kornblith2019similarity}
Simon Kornblith, Mohammad Norouzi, Honglak Lee, and Geoffrey Hinton. 2019.
\newblock Similarity of neural network representations revisited.
\newblock In \emph{International Conference on Machine Learning}, pages
  3519--3529. PMLR.

\bibitem[{Kudo(2018)}]{kudo-2018-subword}
Taku Kudo. 2018.
\newblock \href {https://doi.org/10.18653/v1/P18-1007} {Subword regularization:
  Improving neural network translation models with multiple subword
  candidates}.
\newblock In \emph{Proceedings of the 56th Annual Meeting of the Association
  for Computational Linguistics (Volume 1: Long Papers)}, pages 66--75,
  Melbourne, Australia. Association for Computational Linguistics.

\bibitem[{Kudo and Richardson(2018)}]{kudo-richardson-2018-sentencepiece}
Taku Kudo and John Richardson. 2018.
\newblock \href {https://doi.org/10.18653/v1/D18-2012} {{S}entence{P}iece: A
  simple and language independent subword tokenizer and detokenizer for neural
  text processing}.
\newblock In \emph{Proceedings of the 2018 Conference on Empirical Methods in
  Natural Language Processing: System Demonstrations}, pages 66--71, Brussels,
  Belgium. Association for Computational Linguistics.

\bibitem[{Kurita et~al.(2019)Kurita, Vyas, Pareek, Black, and
  Tsvetkov}]{kurita-etal-2019-measuring}
Keita Kurita, Nidhi Vyas, Ayush Pareek, Alan~W Black, and Yulia Tsvetkov. 2019.
\newblock \href {https://doi.org/10.18653/v1/W19-3823} {Measuring bias in
  contextualized word representations}.
\newblock In \emph{Proceedings of the First Workshop on Gender Bias in Natural
  Language Processing}, pages 166--172, Florence, Italy. Association for
  Computational Linguistics.

\bibitem[{Larson(2017)}]{larson2017gender}
Brian Larson. 2017.
\newblock Gender as a variable in natural-language processing: Ethical
  considerations.
\newblock In \emph{Proceedings of the First ACL Workshop on Ethics in Natural
  Language Processing}, pages 1--11.

\bibitem[{May et~al.(2019)May, Wang, Bordia, Bowman, and
  Rudinger}]{may-etal-2019-measuring}
Chandler May, Alex Wang, Shikha Bordia, Samuel~R. Bowman, and Rachel Rudinger.
  2019.
\newblock \href {https://doi.org/10.18653/v1/N19-1063} {On measuring social
  biases in sentence encoders}.
\newblock In \emph{Proceedings of the 2019 Conference of the North {A}merican
  Chapter of the Association for Computational Linguistics: Human Language
  Technologies, Volume 1 (Long and Short Papers)}, pages 622--628, Minneapolis,
  Minnesota. Association for Computational Linguistics.

\bibitem[{Morcos et~al.(2018)Morcos, Raghu, and Bengio}]{NEURIPS2018_a7a3d70c}
Ari Morcos, Maithra Raghu, and Samy Bengio. 2018.
\newblock \href
  {https://proceedings.neurips.cc/paper/2018/file/a7a3d70c6d17a73140918996d03c014f-Paper.pdf}
  {Insights on representational similarity in neural networks with canonical
  correlation}.
\newblock In \emph{Advances in Neural Information Processing Systems},
  volume~31. Curran Associates, Inc.

\bibitem[{Mu and Viswanath(2018)}]{mu2018allbutthetop}
Jiaqi Mu and Pramod Viswanath. 2018.
\newblock \href {https://openreview.net/forum?id=HkuGJ3kCb} {All-but-the-top:
  Simple and effective postprocessing for word representations}.
\newblock In \emph{International Conference on Learning Representations}.

\bibitem[{Nayak(2019)}]{nayak_2019}
Pandu Nayak. 2019.
\newblock \href
  {https://blog.google/products/search/search-language-understanding-bert/}
  {Understanding searches better than ever before}.

\bibitem[{Ott et~al.(2018)Ott, Auli, Grangier, and Ranzato}]{ott2018analyzing}
Myle Ott, Michael Auli, David Grangier, and Marc'Aurelio Ranzato. 2018.
\newblock Analyzing uncertainty in neural machine translation.
\newblock In \emph{International Conference on Machine Learning}.

\bibitem[{Pennington et~al.(2014)Pennington, Socher, and
  Manning}]{pennington-etal-2014-glove}
Jeffrey Pennington, Richard Socher, and Christopher Manning. 2014.
\newblock \href {https://doi.org/10.3115/v1/D14-1162} {{G}lo{V}e: Global
  vectors for word representation}.
\newblock In \emph{Proceedings of the 2014 Conference on Empirical Methods in
  Natural Language Processing ({EMNLP})}, pages 1532--1543, Doha, Qatar.
  Association for Computational Linguistics.

\bibitem[{Peters et~al.(2018)Peters, Neumann, Iyyer, Gardner, Clark, Lee, and
  Zettlemoyer}]{peters-etal-2018-deep}
Matthew Peters, Mark Neumann, Mohit Iyyer, Matt Gardner, Christopher Clark,
  Kenton Lee, and Luke Zettlemoyer. 2018.
\newblock \href {https://doi.org/10.18653/v1/N18-1202} {Deep contextualized
  word representations}.
\newblock In \emph{Proceedings of the 2018 Conference of the North {A}merican
  Chapter of the Association for Computational Linguistics: Human Language
  Technologies, Volume 1 (Long Papers)}, pages 2227--2237, New Orleans,
  Louisiana. Association for Computational Linguistics.

\bibitem[{Provilkov et~al.(2020)Provilkov, Emelianenko, and
  Voita}]{provilkov-etal-2020-bpe}
Ivan Provilkov, Dmitrii Emelianenko, and Elena Voita. 2020.
\newblock \href {https://www.aclweb.org/anthology/2020.acl-main.170}
  {{BPE}-dropout: Simple and effective subword regularization}.
\newblock In \emph{Proceedings of the 58th Annual Meeting of the Association
  for Computational Linguistics}, pages 1882--1892, Online. Association for
  Computational Linguistics.

\bibitem[{Radford et~al.(2019)Radford, Wu, Child, Luan, Amodei, and
  Sutskever}]{radford2019language}
Alec Radford, Jeffrey Wu, Rewon Child, David Luan, Dario Amodei, and Ilya
  Sutskever. 2019.
\newblock Language models are unsupervised multitask learners.
\newblock \emph{OpenAI blog}, 1(8):9.

\bibitem[{Raffel et~al.(2020)Raffel, Shazeer, Roberts, Lee, Narang, Matena,
  Zhou, Li, and Liu}]{JMLR:v21:20-074}
Colin Raffel, Noam Shazeer, Adam Roberts, Katherine Lee, Sharan Narang, Michael
  Matena, Yanqi Zhou, Wei Li, and Peter~J. Liu. 2020.
\newblock \href {http://jmlr.org/papers/v21/20-074.html} {Exploring the limits
  of transfer learning with a unified text-to-text transformer}.
\newblock \emph{Journal of Machine Learning Research}, 21(140):1--67.

\bibitem[{Schuster and Nakajima(2012)}]{schuster2012japanese}
Mike Schuster and Kaisuke Nakajima. 2012.
\newblock Japanese and korean voice search.
\newblock In \emph{2012 IEEE International Conference on Acoustics, Speech and
  Signal Processing (ICASSP)}, pages 5149--5152. IEEE.

\bibitem[{Sen and Wasow(2016)}]{sen2016race}
Maya Sen and Omar Wasow. 2016.
\newblock Race as a bundle of sticks: Designs that estimate effects of
  seemingly immutable characteristics.
\newblock \emph{Annual Review of Political Science}, 19:499--522.

\bibitem[{Sennrich et~al.(2016)Sennrich, Haddow, and
  Birch}]{sennrich-etal-2016-neural}
Rico Sennrich, Barry Haddow, and Alexandra Birch. 2016.
\newblock \href {https://doi.org/10.18653/v1/P16-1162} {Neural machine
  translation of rare words with subword units}.
\newblock In \emph{Proceedings of the 54th Annual Meeting of the Association
  for Computational Linguistics (Volume 1: Long Papers)}, pages 1715--1725,
  Berlin, Germany. Association for Computational Linguistics.

\bibitem[{Tenney et~al.(2019)Tenney, Das, and
  Pavlick}]{DBLP:journals/corr/abs-1905-05950}
Ian Tenney, Dipanjan Das, and Ellie Pavlick. 2019.
\newblock \href {http://arxiv.org/abs/1905.05950} {{BERT} rediscovers the
  classical {NLP} pipeline}.
\newblock \emph{CoRR}, abs/1905.05950.

\bibitem[{Toney-Wails and Caliskan(2021)}]{toney2020valnorm}
Autumn Toney-Wails and Aylin Caliskan. 2021.
\newblock Valnorm quantifies semantics to reveal consistent valence biases
  across languages and over centuries.
\newblock \emph{Empirical Methods in Natural Language Processing (EMNLP)}.

\bibitem[{Tzioumis(2018)}]{tzioumis2018demographic}
Konstantinos Tzioumis. 2018.
\newblock Demographic aspects of first names.
\newblock \emph{Scientific data}, 5(1):1--9.

\bibitem[{Voita et~al.(2019)Voita, Sennrich, and
  Titov}]{voita-etal-2019-bottom}
Elena Voita, Rico Sennrich, and Ivan Titov. 2019.
\newblock \href {https://doi.org/10.18653/v1/D19-1448} {The bottom-up evolution
  of representations in the transformer: A study with machine translation and
  language modeling objectives}.
\newblock In \emph{Proceedings of the 2019 Conference on Empirical Methods in
  Natural Language Processing and the 9th International Joint Conference on
  Natural Language Processing (EMNLP-IJCNLP)}, pages 4396--4406, Hong Kong,
  China. Association for Computational Linguistics.

\bibitem[{Wang et~al.(2020)Wang, Lin, Rajani, McCann, Ordonez, and
  Xiong}]{wang2020doublehard}
Tianlu Wang, Xi~Victoria Lin, Nazneen~Fatema Rajani, Bryan McCann, Vicente
  Ordonez, and Caiming Xiong. 2020.
\newblock Double-hard debias: Tailoring word embeddings for gender bias
  mitigation.
\newblock In \emph{Association for Computational Linguistics (ACL)}.

\bibitem[{Wendlandt et~al.(2018)Wendlandt, Kummerfeld, and
  Mihalcea}]{wendlandt-etal-2018-factors}
Laura Wendlandt, Jonathan~K. Kummerfeld, and Rada Mihalcea. 2018.
\newblock \href {https://doi.org/10.18653/v1/N18-1190} {Factors influencing the
  surprising instability of word embeddings}.
\newblock In \emph{Proceedings of the 2018 Conference of the North {A}merican
  Chapter of the Association for Computational Linguistics: Human Language
  Technologies, Volume 1 (Long Papers)}, pages 2092--2102, New Orleans,
  Louisiana. Association for Computational Linguistics.

\bibitem[{Wolf et~al.(2020)Wolf, Debut, Sanh, Chaumond, Delangue, Moi, Cistac,
  Rault, Louf, Funtowicz, Davison, Shleifer, von Platen, Ma, Jernite, Plu, Xu,
  Scao, Gugger, Drame, Lhoest, and Rush}]{wolf-etal-2020-transformers}
Thomas Wolf, Lysandre Debut, Victor Sanh, Julien Chaumond, Clement Delangue,
  Anthony Moi, Pierric Cistac, Tim Rault, Rémi Louf, Morgan Funtowicz, Joe
  Davison, Sam Shleifer, Patrick von Platen, Clara Ma, Yacine Jernite, Julien
  Plu, Canwen Xu, Teven~Le Scao, Sylvain Gugger, Mariama Drame, Quentin Lhoest,
  and Alexander~M. Rush. 2020.
\newblock \href {https://www.aclweb.org/anthology/2020.emnlp-demos.6}
  {Transformers: State-of-the-art natural language processing}.
\newblock In \emph{Proceedings of the 2020 Conference on Empirical Methods in
  Natural Language Processing: System Demonstrations}, pages 38--45, Online.
  Association for Computational Linguistics.

\bibitem[{Yang et~al.(2019)Yang, Dai, Yang, Carbonell, Salakhutdinov, and
  Le}]{NEURIPS2019_dc6a7e65}
Zhilin Yang, Zihang Dai, Yiming Yang, Jaime Carbonell, Russ~R Salakhutdinov,
  and Quoc~V Le. 2019.
\newblock \href
  {https://proceedings.neurips.cc/paper/2019/file/dc6a7e655d7e5840e66733e9ee67cc69-Paper.pdf}
  {Xlnet: Generalized autoregressive pretraining for language understanding}.
\newblock In \emph{Advances in Neural Information Processing Systems},
  volume~32. Curran Associates, Inc.

\bibitem[{Zhou et~al.(2021)Zhou, Ethayarajh, and
  Jurafsky}]{zhou2021frequencybased}
Kaitlyn Zhou, Kawin Ethayarajh, and Dan Jurafsky. 2021.
\newblock \href {http://arxiv.org/abs/2104.08465} {Frequency-based distortions
  in contextualized word embeddings}.

\bibitem[{Zhu et~al.(2015)Zhu, Kiros, Zemel, Salakhutdinov, Urtasun, Torralba,
  and Fidler}]{zhu2015aligning}
Yukun Zhu, Ryan Kiros, Rich Zemel, Ruslan Salakhutdinov, Raquel Urtasun,
  Antonio Torralba, and Sanja Fidler. 2015.
\newblock Aligning books and movies: Towards story-like visual explanations by
  watching movies and reading books.
\newblock In \emph{Proceedings of the IEEE international conference on computer
  vision}, pages 19--27.

\end{thebibliography}
\bibliographystyle{acl_natbib}
\clearpage
\appendix
\section{Neural Language Models}
\label{nlm_appendix}
\subsection{BERT}

\citet{devlin-etal-2019-bert} introduce BERT, a bidirectional transformer trained on masked language modeling (also known as the  \textit{Cloze} objective) and next sentence prediction tasks. We examine the “bert-base-cased” model available from Hugging Face Transformers. BERT is trained on 2.5 billion words of English Wikipedia documents (excluding lists, headers, and tables) and 800 million words of the BookCorpus compiled by \citet{zhu2015aligning}, a collection of more than 11,000 free online books written by unpublished authors.

\subsection{GPT-2}

\citet{radford2019language} introduce OpenAI GPT-2, a unidirectional transformer model trained on the next-word prediction language modeling task. We examine the “gpt2” model available from Hugging Face Transformers. GPT-2 is trained on the WebText corpus, a web scrape composed of outbound links from Reddit with at least 3 karma, which the model designers (who are also the creators of the training corpus) take as a heuristic for link quality. WebText contains over 8 million documents and 40GB of text.

\subsection{XLNet}

\citet{NEURIPS2019_dc6a7e65} introduce XLNet, an autoregressive transformer trained on language modeling which learns bidirectional contexts by maximizing the expected likelihood over all permutations of its input’s factorization order. We use the “xlnet-base-cased” model available from Hugging Face Transformers. Like BERT, XLNet is trained on English Wikipedia and the BooksCorpus. However, it also trains on three additional corpora: Giga5 (a corpus of print news from publications such as the Associated Press and New York Times), ClueWeb 2012-B, and Common Crawl. ClueWeb 2012-B is a corpus maintained by Carnegie Mellon University.

\subsection{T5}
\citet{JMLR:v21:20-074} introduce T5, a text-to-text encoder-decoder transformer model trained on a number of different pretraining tasks and designed to take in text and return text for any NLP task. T5 is pretrained on the Colossal Cleaned Crawled Corpus (C4). We examine the encoder layers of the 't5-base' model available on Hugging Face Transformers.
\section{Subword Tokenizers}
\label{tokenizer_appendix}
\subsection{Byte-Pair Encoding}

GPT-2 uses a variation on Byte-Pair Encoding (BPE) to iteratively choose the most frequently occurring bigram of symbols in the training corpus, merge them into a single symbol, and add the merged symbol to its subword vocabulary until it reaches its maximum vocabulary size of 50,256. BPE is a compression algorithm originally proposed by \citet{gage1994new}. The technique was adapted for encoding rare subwords for the purpose of neural machine translation in 2015 by \citet{sennrich-etal-2016-neural}.  GPT-2 modifies this algorithm to operate directly on byte sequences rather than Unicode character sequences.

\subsection{WordPiece}

BERT uses the WordPiece tokenization algorithm, first developed by \citet{schuster2012japanese} and subsequently refined by \cite{johnson2017google} in Google's neural machine translation system. WordPiece builds a probability distribution from the training corpus using the base vocabulary of single character symbols, and then iteratively assembles its vocabulary by merging a bigram, adding the merged unit, and replacing every instance of the bigram in the training corpus with the merged unit. WordPiece selects the bigram which most increases the likelihood of the training data when the merged bigram is added to the language model, a process which is indirectly related to frequency.

\subsection{SentencePiece}

XLNet and T5 use the SentencePiece tokenizer, which can implement either Byte-Pair Encoding or Unigram to form a vocabulary. SentencePiece was developed by  \citet{kudo-richardson-2018-sentencepiece} and allows for lossless tokenization, such that original input can be reconstructed from tokenized input. SentencePiece inserts an underscore to preserve whitespace as a character, and can be used with byte-pair encoding or with the Unigram algorithm also introduced by \citet{kudo-2018-subword}. Unlike WordPiece and BPE, Unigram starts with a large set of words and subwords generated from a training corpus, and iteratively removes the words which have the least effect on the overall loss (with the loss function chosen as a hyperparameter) of a language model created from the vocabulary until a fixed vocabulary size is reached.
\section{First Names Dataset}
\label{first_names_appendix}
The names dataset of \citet{tzioumis2018demographic} provides the percentage of individuals with a first name who self-identify as each of six races and ethnicities: White, Black, Hispanic, Asian-Pacific-Islander, Native American, and Mixed Race. We provide information regarding the number of names in each group after cross-referencing for gender with the 1990 U.S. Social Security Administration data in Table \ref{Dataset_Intersectional_Names_Table}. Note that we were left with only one predominantly Native American name and one predominantly Mixed Race name, which is not enough names for use in our experiments.

\begin{table}[h]
\centering
\begin{tabular}
{|l||c|}
 \hline
 \multicolumn{2}{|c|}{Names by Intersectional Group} \\
 \hline

Intersectional Group & Number of Names\\
 \hline
   Asian Female & 156\\
 Black Female & 37\\
Hispanic Female & 189\\
White Female & 1,621\\
   Asian Male & 263\\
Black Male & 32\\
Hispanic Male & 243\\
White Male & 1,216\\
 \hline
\end{tabular}
\caption{Number of names in our dataset}
\label{Dataset_Intersectional_Names_Table}
\end{table}

After cross-referencing between Tzioumis’ list and the 1990 U.S. Social Security Administration data, we are left with a list of 3,757 first names. White-majority names are more common in this list, likely due both to the United States being a plurality-white nation, and to economic and structural disparities which allow easier access to mortgages for white individuals.

\end{document}